# A Web Infrastructure for Certifying Multimedia News Content for Fake News Defense

Edward L. Amoruso
Department of Electrical
Computer Engineering
University of Central Florida
Orlando, FL 32816 US
eamoruso@knights.ucf.edu

Stephen P. Johnson
Department of Computer Science
University of Central Florida
Orlando, FL 32816 US
johnsons@knights.ucf.edu

Raghu Avula
Department of Electrical and
Computer Science
University of Central Florida
Orlando, FL 32816 US
raghunandan@knights.ucf.edu

Cliff C. Zou
Department of Computer Science
University of Central Florida
Orlando, FL 32816 US
changchun.zou@cs.ucf.edu

***Abstract*— In dealing with altered multimedia news content, also referred to as fake news, we present a ready-to-deploy scheme based on existing public key infrastructure as a new fake news defense paradigm. This scheme enables news organizations to certify/endorse a newsworthy multimedia news content and securely and conveniently pass this trust information to end users. A news organization can use our program to digitally sign the multimedia news content with its private key. By installing a browser extension, an end user can easily verify whether a news content has been endorsed and by which organization. It is totally up to the end user whether to trust the news or the endorsing news organization. The underlining principles of our scheme are that fake news will sooner or later be identified as fake by general population, and a news organization puts its long-term reputation on the line when endorsing a news content.**

## I. Introduction

This paper was motivated by the difficulty associated with identifying fake news, particularly the misleading visual multimedia content that deludes users on social networks. The use of images and videos on the internet have been a driving force in forging opinions. With the plethora of new software tools available to the common end-user, it has become effortless to produce fake images and misrepresenting as valid content. These fake images can look plausible enough making them more dangerous with social media's ability to spread them across the globe before they can be debunked by fact-checking sites. It is estimated that over three billion images are shared every day on social media and three hundred hours of video per minute are uploaded to YouTube [1]. With such statistics and aggressive competition of news organizations trying to publish their story first, journalists are forced to spend less time validating the provenance of images and videos.

Research has shown that visual content can affect public opinion, leading to various levels of confusion [2]. One example of altered visual multimedia was seen when someone published a fake photo of Mitt Romney, shown in Fig. 1, presidential candidate, spelling the word "Money" instead of his last name "Romney". Fig. 1(b) was posted on Facebook adding bias and negative opinion with the following caption, "The Romney family misspells their own name in what might be the greatest Freudian slip in US history" [3]. This post also received over three thousand comments and was shared to over ninety thousand users just on Facebook alone. A study performed in 2018 conveyed that Facebook was considered the primary platform for social media. It concluded that 68% of U.S adults used Facebook, with three-quarters accessing the site daily [4]. According to a survey by Pew Research Center performed in December of 2016, about 64% of the 1,002 U.S. adult citizens interviewed said fabricated news stories cause a great deal of confusion about the basic facts of current issues [2].

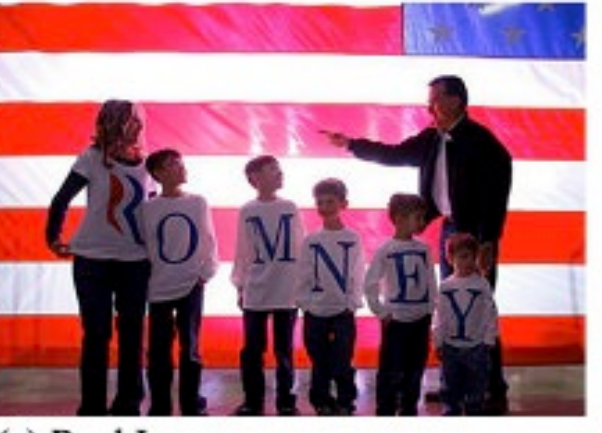

(a) Real Image

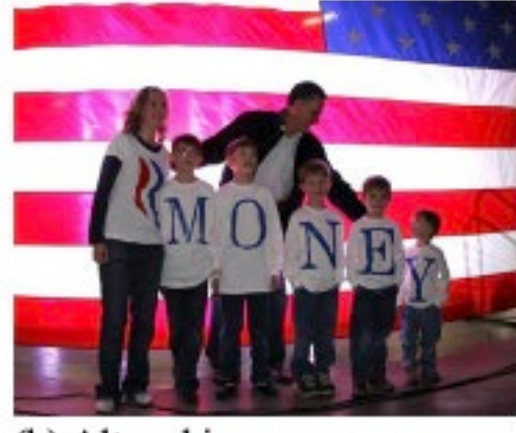

(b) Altered image

Fig. 1. Example showing a real image (a) and fake image (b) of Mitt Romney posing for a picture prior to a presidential rally.

However, fighting fake news is very hard and there are many technical and even political challenges. Fake news publishers can use sophisticated techniques to modify an image/video without leaving the trace, use real image/video taken in older time, or use real multimedia content captured at a different location. In addition, there often exists some political force pushing fake news to general audience.

Therefore, in this paper we are not proposing another technical approach to verify multimedia content's integrity or authority like much prior research; instead, our proposed system enables and encourages a traditional news organization to conveniently endorse a news content when it believes the news content is real. On the news receiving end, our developed browser extension enables an end user to easily identify which news multimedia content is certified and by which news organization. We want to emphasize that it is totally up to the user whether to trust the news or the endorsing news organization.

Fake news has a unique feature as it is time-sensitive, and sooner or later will be discovered to be fake by the general audience. Therefore, a news organization is incentivized to do endorsement and only endorse true news since its reputation is on the line and organization's reputation is long-lasting and critical to its survival. This will guarantee that our developed platform can be practically and widely adopted, and all well-known news organizations will not abuse the platform. If some organizations abuse the system by endorsing fake news, few audiences will believe the endorsement since those organizations are either unknown or have bad reputation. The endorsing/certification of a news content by a news organization puts the news organization's long-term reputation on the line, which will prevent irresponsive endorsement or abuse of our proposed platform.

The contributions of this paper are to provide:

- A new paradigm on fighting against fake news by providing a trusted communication channel between traditional news organizations and end users, instead of directly verifying the truthiness of a news piece by the proposed system.
- An easy to implement infrastructure for news organizations to endorse/certify the truthiness of a news piece and enables an end user to easily verify the endorsement of a news piece and by which news organization via our developed browser extension.

The rest of this paper is organized as follows. Section II covers an overview of other related works and how our approach differs. In Section III, we introduce our proposed approach and cover the design of both the server and client methodology. Section IV we discuss the implementation of both the server file authentication and endorsement, and the client browser extension. Section V will go over both the limitations and future work of our paper. Finally, Section VI provides our conclusion.

## II. Related Work

### A. *Fake News Defense in Political and Social Science*

Fake news with altered visual multimedia has become a common tactic in confusing and deceiving readers on the Internet for either political or personal gain. In scenarios of deception, the user's interpretation can cause a reaction of panic and further proliferation of the fake visual multimedia content. More dangerously, the users may experience fear or ire leading to, for example, the selling of stocks or even creating riots [5]. Other motivations for creating fake visual multimedia content are to lure users to a specific website, also referred to as "clickbait's", for advertisement revenue gain [5].

In response, there have been several approaches proposed by different entities to help detect fake visual multimedia content. Social media companies such as Google, Facebook, and Twitter have attempted to solve this problem by denying individuals, associated with the publishing of misleading information, from acquiring revenue from clicks and increased traffic [6]. Under the assumption that credible users provide credible tweets, users on Twitter users that spread fake news are credible users who are unable to verify the news and spread it unintentionally [7]. These companies have contributed little to solving the spread of fake visual multimedia content [6]. Other research has proposed solutions that can detect and filter out websites containing false and misleading information. These approaches typically require a custom tool to be downloaded and installed by the end user. Aldwairi et al. presented such a tool in identifying and blocking fake news. According to the authors of this paper, once the tool is enabled, it uses various techniques such as syntactic characteristics to detect misinformation. Other proposed works involve image analysis techniques on visual multimedia content to determine their validity [8], [9], [10]. These solutions, however, introduce complexity that could easily deter news agencies from utilizing them.

Tagging or watermarking techniques have been presented for digital right management and multimedia content ownership tracing [11]. In fake news detection we aim at detecting content manipulation, which is different from what watermarking tries to achieve.

### B. *Related Technological Approaches in Fake News Defense*

During our research of proposed works that could detect and warn end users of fake multimedia content, we were unable to find any that used our approach. We searched for detection techniques that did not involve the modification of the original visual multimedia content. Other emphasis in our search was placed on techniques that utilized the CA's (Certificate Authority) private and public keys for use in visual multimedia verification process.

The closely related work is presented in paper [12] where the authors used a similar component to footprint an image for validation. They only focus on embedding the metadata within the image file. They state, "Inserting the certificate into the file is the only apparent reliable method and is the approach used by leading companies such as Microsoft and Adobe for digital signing of Office and PDF documents" [12]. To achieve image verification, Harran, Farrelly, and Curran impose the use of an accepted standard called XMP (Extensible Metadata Platform) to add metadata to various formats of image files, especially JPEG's. XMP essentially provides a roadmap for metadata placement in several types of image files [13].

### C. *How Our Approach Differs from Other Approaches*

In our approach, the image or video file is kept in its original format, opposed to other related works. In doing so, we can maintain the multimedia visual content integrity. This also plays an important role as you will see later in our design. We take advantage of XMP (Extensible Metadata Platform) without the modification of any multimedia visual files, eliminating third party software altogether, by solely using the XMP sidecar file for establishing a digital footprint. The digital footprint stores both the visual multimedia file's hash and digital certificate for use in the validation process by our web browser extension. This approach eliminates the embedding of metadata within the image file, removing overhead and complexity, for establishing visual multimedia validation. Furthermore, in our approach the certifier can add additional information associated with the multimedia file such as its timestamp, geolocation, brief description of the story, etc., which gives the certifier more freedom and control on what they want to endorse.

## III. Our Proposed Approach

### A. Objective and Design

The proposed approach is one that maintains the integrity of the image or video file without any modification while at the same time providing a hash and digital signature for validation and verification. When a news organization decides to endorse/certify a multimedia news content, it can use our proposed platform to process the content to generate cryptographic hash and digital signature, which is then embedded into the generated XMP file located together with the image/video file. Our platform will then modify the news HTML page to contain a metatag which points to the XMP sidecar file for that image. These changes can be seen in Table II and III as the before and after.

Our design can be used on any forms of multimedia news content (text, image, video, etc.). For explanation purpose, in this paper we will mainly use image as the example for our design description.

An end user can install a browser extension to display those images/videos with visual information indicating the verification results. An example of the colored border around a verified/failed image content can be seen in Fig. 2. For verified image, besides the added, green-colored border, the browser extension will also display the authenticator's name on the image so that the end user can judge whether to trust this certification or not.

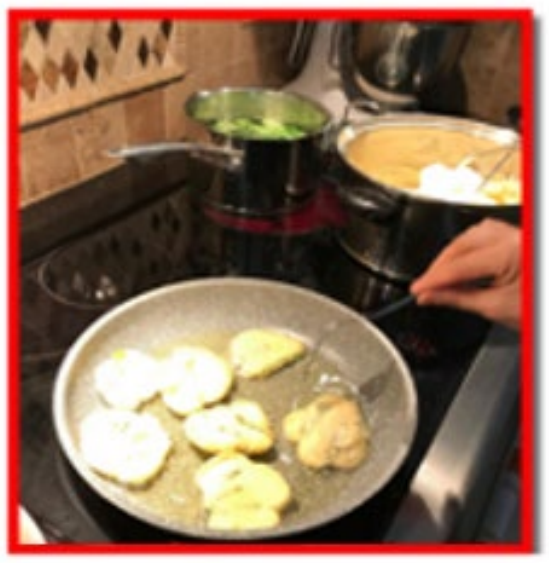

(a) Image Failed Verification

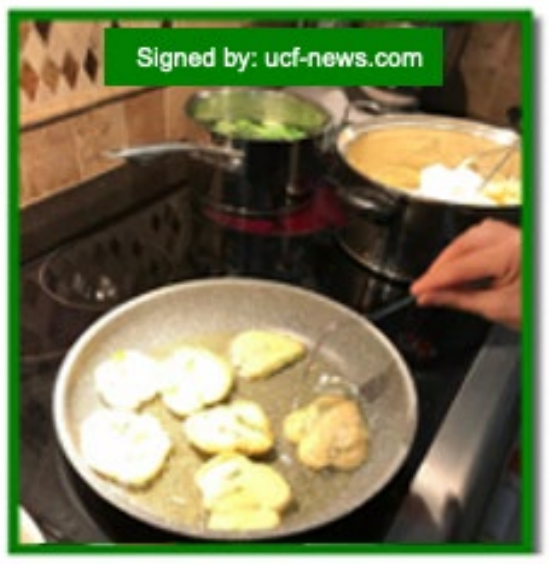

(b) Image Passed Verification

Fig. 2. Browser Extension notification for each image in a webpage. A colored border is displayed around the image after the browser extension performs validation. When the browser extension detects that an image has been modified, a red border is displayed with the image as shown in (a). A green border is displayed when the browser extension successfully validates the image as shown in (b); it also displays the authenticator's name so that the end user can judge whether to trust this certification or not.

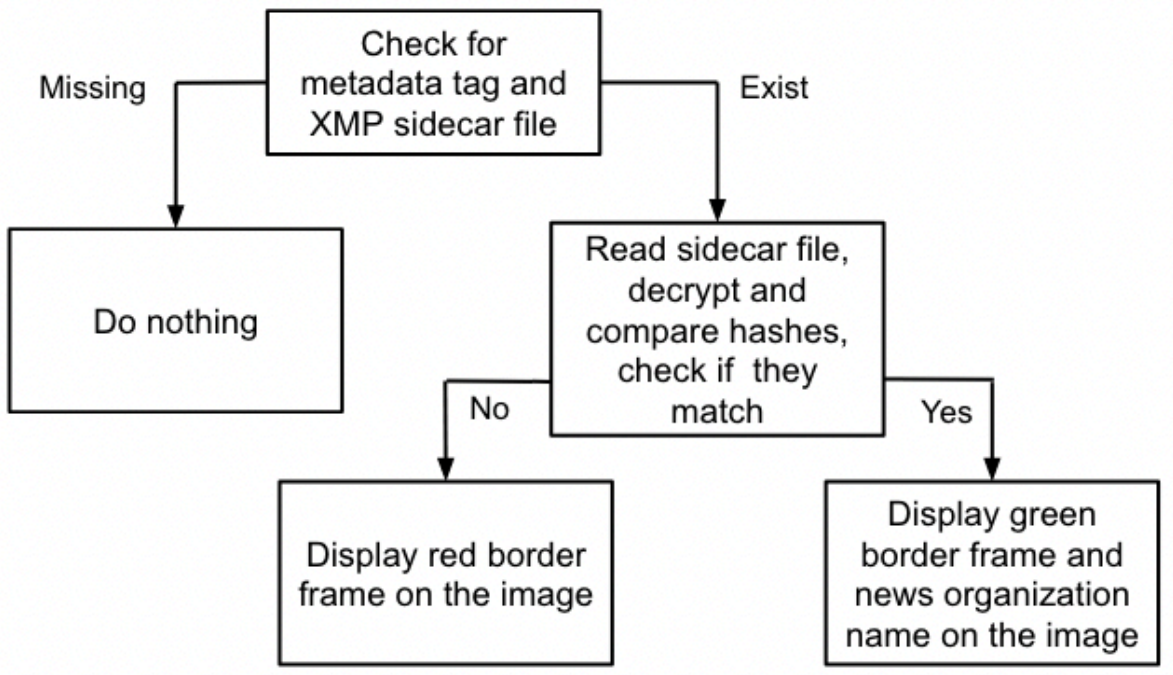

Fig. 3. Flow-chart of the browser extension processing each image in a loaded webpage.

Fig. 3 demonstrates the browser extension's process flow as the webpage is loaded by a client's web browser. Our browser extension will first search for the metatag identified with "x-media-cert", example shown in Table III, and then proceed by reading the XMP sidecar file representing the image. On completion, it presents the client a green or red border around the image to indicate its verification result. If the "x-media-cert" metatag is not found, or sidecar file is missing, the browser extension will ignore the image and display it in its original form, making it transparent to the end user's web browser.

Our designed platform mainly contains two parts. One is the program utilized by news organizations to certify a news multimedia content; the other is the browser extension program that can be installed and used by end users to verify the certified news content on webpages.

In summary, the design goals of our proposed platform are:

- We do not provide news authentication or verification by our platform itself, but instead provide a platform that facilitate news organizations to certify/endorse a multimedia news content they believe is truthiness and newsworthy.
- End user-side component is realized by a browser extension so that it can be easily installed by users who want to use this proposed fake news defense service.
- Our designed platform is transparent to end users who are not using our system and does not affect their browsing experience. The added certification information by news organization is ignored by the end user's web browser if the browser extension is disabled or not installed.
- Multimedia news content can be easily forwarded or reposted by other web publishers and still maintain the authentication information.

### B. Extensible Metadata Platform (XMP)

Using already developed standards provides leverage for seamless integration with various platforms and technologies. In our design, XMP (Extensible Metadata Platform) was applied to help store metadata in XML (Extensible Markup Language) format.

XMP is a data model that provides a unique name and value to broadly represent data types, especially those used in digital file formats. XMP's design provides the roadmap for embedding metadata information into an image, video, and document file formats. Metadata describes the properties of a resource, for example, in a jpeg image, the content will include actual binary data while the metadata contains properties such as author, creation date, location, and other information about the image. XMP sidecar files are not allowed to contain any binary data, but instead use various encoding methods (e.g., *base64*) to allow the inserting of metadata information [13].

### C. Digital Signature using Public Key Infrastructure (PKI)

The digital signature, also referred to as the signed message digest, is used to digitally sign a message ($m$). The XMP file contains both the digital signature and the hash values. It also contains the essential metadata of the news image content, such as the author, picture taken time, geolocation, and a brief text description. We want to emphasize that the digital signature is not the hash value H(m), it is the encrypted H(m) using the Authenticator's private key. The private key used is originally obtained by a trusted Certificate Authority (CA), who verifies the identity of who you are communicating with (e.g., Authenticator) online.

The receiver needs to extract the public key of the Authenticator (from the Authenticator's digital certificate), decrypt the digital signature by using the Authenticator's public key, and then compare the results (hash value) with the hash value computed by itself to see if they match or not. In Table I we provide a list of notations used in our paper.

TABLE I. MAIN NOTATIONS USED IN THIS PAPER.

| $K_A^+, K_A^-$ | Authenticator's public key ($^+$) and private key ($^-$) |
|---|---|
| $K_{CA}^+, K_{CA}^-$ | Certificate Authority's public key ($^+$) and private key ($^-$) used in our certification procedure |
| $K_{CA}^-(K_A^+)$ | **Digital certificate** for the authenticator's public key $K_A^+$ |
| *Metadata* | The essential information about an image added by the authenticator, including date, geolocation, brief description of the image story |
| *m* | Digital content of an image and its Metadata |
| *H(m)* | Hash value of data content *m*, e.g., SHA256 digest of the content *m* |
| $K_A^-(H(m))$ | **Digital signature** of the data content *m* signed by the Authenticator |

## D. Server Side Design

The server-side program refers to our developed program utilized by news organizations to endorse/certify a multimedia news content. The server's implementation is typically performed by the news agency's web development team. These developers will require some working knowledge of how to manipulate image tags in the HTML file containing the multimedia news content. Once the news agency has verified the visual multimedia content it is up to them to use our server-side program to create the metadata and XMP sidecar file and include it alongside with the image file. In our research, scripts and other tools have been developed to assist the developers with the creation of the XMP sidecar file. Prior to generating the digital signature, the news agency must have acquired an RSA private/public key pair for the authenticator's website.

Below we break down each step involved in processing an image and generating the associated XML file, also referred to as the XMP sidecar file, by the authenticator:

- Authenticator's personnel inputs Metadata for the image (author, date, and geolocation of the image, and a brief description text of the image story)
- **Concatenate** the image content with the metadata together (for example, as a temporary file)
- Create SHA256 digest, i.e., $H(Metadata + image)$.
- Create the **digital signature** we need, i.e., $K_A^-(H(Metadata + image))$.
- Create the image's **associated XMP** file, containing: Metadata + $K_A^-(H(Metadata + image))$ + $K_{CA}^-(K_A^+)$
- Modify the webpage snippet containing the image reference (see example in Table II) to add the reference to the associated XML file (see example in Table III).

Both the visual multimedia and XMP sidecar files must be place in the same directory on the webserver. These files have the same name except for their file extensions which identify them as either the visual multimedia file (e.g., filename.jpg) or the XMP sidecar file (e.g., *filename.xml*). Finally, the HTML page containing the reference to the visual multimedia file, such as the snippet shown in Table II, is then updated by our server-side program by adding additional values to the image tag in the markup language with the reference of the associated XMP sidecar file as shown in Table III.

TABLE II. EXAMPLE OF AN ORIGINAL HTML PAGE SNIPPET CONTAINING THE REFERENCE TO AN IMAGE

| <img src="/images/img5914.jpeg" id = "img_12e144f2" alt = "Cooking at home" title = "Ceramic Cookware"> |
|---|

TABLE III. MODIFIED HTML PAGE SNIPPET ADDING ATTRIBUTES TO THE IMAGE TAG. NOTE THE HIGHLIGHTED CODE POINTS TO THE XMP SIDECAR FILE (AKA XML) THAT IS USED BY THE BROWSER EXTENSION TO VALIDATE THE IMAGE

| <img src="/images/img5914.jpeg" id = "img_12e144f2" alt = "Cooking at home" title = "Ceramic Cookware" x-media-cert="/images/img5914.jpeg.xml"> |
|---|

## E. Client Side Design

The client side is unique in that it only requires a browser extension to accomplish its verification of the visual multimedia content. The browser extension, which is written in pure JavaScript, can be downloaded from the web browsers extension store.

In the following, we provide the client-side browser extension code operation procedure:

- Scan for and load webpage images and associated XML files, also referred to as the XMP sidecar files.
- For Each image that has an associated XML file:
- Extract the digital certificate of the authenticator from XML file, i.e., $K_{CA}^-(K_A^+)$.
- Process this digital certificate to extract: (1). Authenticator's name; (2). Authenticator's public key $K_A^+$.
- Extract the digital signature contained in the XML file, i.e., $K_A^-(H(Metadata + image))$.
- Use $K_A^+$ to decrypt the digital signature to recover and obtain the digest provided by the Authenticator, i.e., $H(Metadata + image)$.
- Extract the Metadata contained in the XML file, concatenate it with the image file.

- Compute the SHA256 digest by the browser extension itself using the same hash function, i.e., $H'(Metadata + image)$.
- Compare the extracted digest provided by the Authenticator, $H(Metadata + image)$, with the browser extension's self-computed digest $H'(Metadata + image)$. If they match, the image and its metadata are verified (adding a green-colored border around the image in our implementation) and continue to process the next image; otherwise, match failed (adding a red-colored border around the image in our implementation) and finish this image operation.
- On verified image, display information showing details of certification. When the user wants to know more about the certification information, a simple mouse over the image will display a green box over it with essential metadata information pertaining to that image. The information includes Authenticator's name, author's name, image taken date, Geolocation of the image, and brief text description of the image story and can be seen in Fig. 4(b).

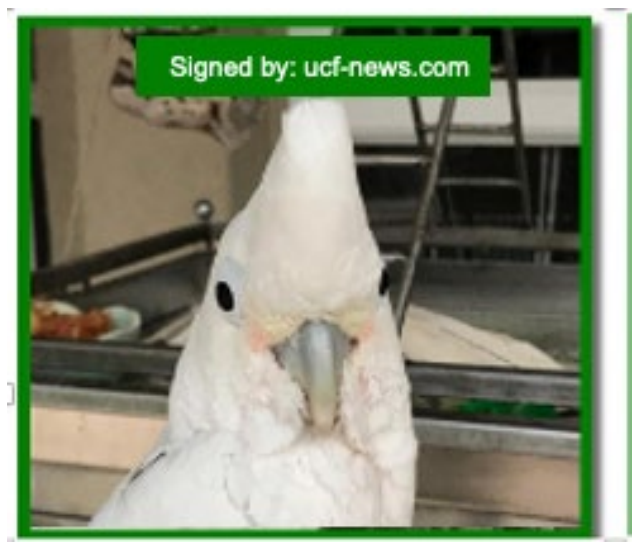

(a) Without mouse hovering over.

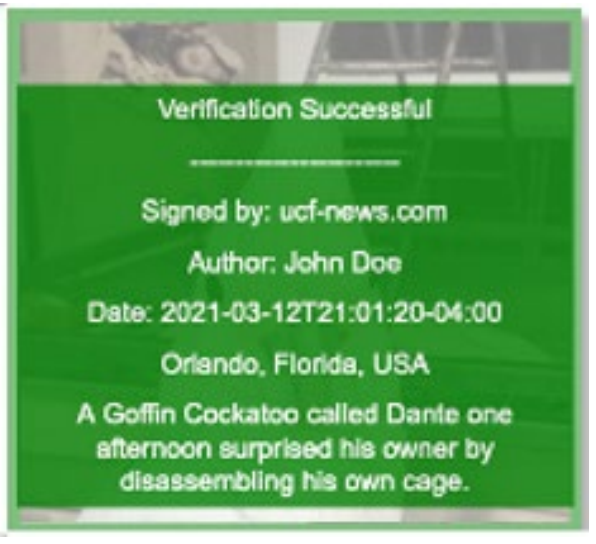

(b) With mouse hovering over

Fig. 4. Display of certification information on an example certified image. Shown in (b), by hovering over the image with the mouse in the green border added by the browser extension, a pop-up window provides the following essential metadata information of the image: Endorsing organization's name, Date & Time, Geolocation, Photographer, and brief text description.

## IV. Implementation

Our implementation is broken up into two parts. The first part, referred as 'server-side implementation', involves the setup and staging of file authentication and endorsement, which should be implemented by news organizations who want to certify some newsworthy visual multimedia contents. The second part of our implementation is the browser extension, which can be easily installed onto the web browser by an end user to process the visual multimedia content and display the certification and endorsement information.

Our implementation's source code and browser extension are publicly available for download on GitHub with following address: https://github.com/eamoruso/FakeNewsDefense.

### A. Server File Authentication and Endorsement

The server-side implementation is platform independent, allowing it to be hosted by various web server technologies such as Microsoft Internet Information Services (IIS), Apache, Nginx, and any others that support the HTTPS protocol. It is important to understand that we only require two things for the server-side to work, the image file and XMP sidecar file, the rest is left to our browser's extension program. Preparing the server file authentication and endorsement can be accomplished manually or automatically by using our shell script program. The shell script program is written in BASH (Bourne Again SHell) and is a Unix/Linux shell command language.

Our automated script is executed in the same directory where the images are located, prompting the user for required inputs such as the date (*dateTimeValue*), geolocation (*cityValue*, *regionValue*, *countryValue*), author (*creatorValue*), and a brief text description of the news story (*headlineValue, descriptionValue*). Once the information is collected, our script then creates the XMP sidecar file based on those inputs. The file created will have the image file name plus an additional extension ".xml" added (e.g., *imagefilename.jpeg.xml*). In the case where our shell script is not supported by the organization's environment (e.g., Microsoft Windows), we provide instructions on our GitHub page for manually creating the XMP sidecar file.

### B. Client-side Browser Extension

The client interface, also referred to as the browser extension, was developed using the JavaScript programming language. Careful design was incorporated to provide a JavaScript implementation that supported browser built-in API (Application Programming Interface) technologies. This would allow the browser extension all the necessary features required to perform its validation and authentication. Another objective was to provide cross platform compatibility with as many operating systems as possible (e.g., Windows, MacOS, and Linux).

To eliminate any dependencies on server-side code and remove the need for additional backend server's or services, we chose to develop our extension on Mozilla's Firefox browser platform. Firefox supports various platforms and contains a rich library of built-in APIs, which allowed us to implement our extension in pure JavaScript code. This approach also provided security benefits as our code doesn't rely on JavaScript Frameworks (e.g., Angular, React) and other libraries such as jQuery.

The browser extension's logic consists of a single loop and two conditions. The loop scans each image in search of an XML file. If it is found, the first condition will process the XML file associated with that image. In the second condition, if the image's digest and signature match, meaning it is valid, it will create a green box around the image with the authenticator's name as shown in Fig. 4a. If the second condition fails, a red box will be placed around the image. If an XML file is not found, the loop will quickly skip the image. This process is continuously repeated until all images on the webpage have been scanned. Finally, after all the images have been processed, the browser extension will enable a mouse hover over feature to display that specific image's information as shown in Fig. 4b.

## V. Evaluation

### A. Server Website and Webpage

For the evaluation we used a public website, hosted on Windows server running Internet Information Services (IIS) version 10.0.14393. The website consisted of a variety of elements such as title, heading, scripts, style sheets and other common meta tags to help mimic a typical news organization's webpage. To test our browser extension's performance, the webpage contained a total of ten valid images.

The time required to manually create the XMP sidecar file for each image averaged three minutes. In this process, we ran our BASH Script and entered the required information for that image (e.g., headline, description, creator's name, date, time, city, region, and country). A news organization could automate this process to reduce the time from our average three minutes to seconds, assuming the images already include metadata information such as headline, description, etc.

### B. Client Software and Hardware

The evaluation's test and results were achieved using Mozilla's Firefox Browser version 89.0.2 in conjunction with our browser extension. Although several computers were used during testing, the performance timings presented here were recorded from an Apple M1 MacBook Pro. This computer consisted of the following hardware specifications: 16GB of memory, 994GB Solid State Drive (SSD), and an eight-core processor.

### C. Evaluation Results

The following results were obtained using ten iterations and taking the average time of those iterations when processing each image with our browser extension (shown in Table IV).

TABLE IV. RESULTS OBTAINED FROM SCANNING TEN IMAGES

| **Image (size)** | **Browser Extension Scans** | | | | | | | | | | |
|---|---|---|---|---|---|---|---|---|---|---|---|
| | ***Iterations*** | | | | | | | | | | ***Avg. Time (ms)*** |
| | ***1*** | ***2*** | ***3*** | ***4*** | ***5*** | ***6*** | ***7*** | ***8*** | ***9*** | ***10*** | |
| IMG1 (21KB) | 28 | 17 | 19 | 18 | 18 | 19 | 27 | 18 | 19 | 28 | **21** |
| IMG2 (28KB) | 22 | 37 | 35 | 44 | 52 | 22 | 21 | 23 | 22 | 24 | **30** |
| IMG3 (37KB) | 40 | 51 | 29 | 29 | 28 | 25 | 29 | 26 | 29 | 29 | **32** |
| IMG4 (41KB) | 30 | 52 | 34 | 36 | 35 | 27 | 36 | 26 | 39 | 29 | **34** |
| IMG5 (43KB) | 84 | 29 | 54 | 30 | 43 | 29 | 30 | 35 | 32 | 30 | **40** |
| IMG6 (45KB) | 37 | 39 | 49 | 33 | 38 | 50 | 35 | 42 | 36 | 36 | **40** |
| IMG7 (160KB) | 115 | 109 | 123 | 101 | 109 | 82 | 98 | 83 | 99 | 85 | **100** |
| IMG8 (163KB) | 102 | 116 | 112 | 118 | 116 | 103 | 106 | 101 | 113 | 100 | **109** |
| IMG9 (190KB) | 114 | 123 | 117 | 119 | 120 | 101 | 113 | 111 | 119 | 115 | **115** |
| IMG10 (214KB) | 124 | 128 | 123 | 129 | 124 | 110 | 149 | 112 | 140 | 121 | **126** |

We only recorded the time from when the logic starts and finishes during image verification. All other variables such as loading the image into the client's browser and retrieving the XMP sidecar file were ignored.

In our test results, we were able to demonstrate that our browser extension used little time, less than quarter of a second, to validate each image showing no visible impact to the users browsing experience. Another factor in our evaluation is that the browser utilizes asynchronous request, allowing the browser extension to perform validation on several images at the same time, thus reducing the overall time. If our browser extension is left enabled during access to websites that do not use our server-side system, there is no impact on performance. When a website is detected to have our server-side system, depending on the number of images and corresponding size, a slight delay is experienced by the end-user as shown in Table IV.

## VI. Limitations and Future Work

### A. Platform Implementation by Social Media System

In cryptography, digital signature guarantees that once a signed message has a single bit change, the resulting message cannot pass the digital signature verification. In terms of our news content certification, it means that a certified news content (such as an image), including its associated metadata (such as organization name, time, geolocation, and brief description), cannot be altered without voiding the certification capability. This feature is problematic when other websites want to forward or repost the news content after some legitimate changes, especially when the certified multimedia news content has large-size images or videos. Specifically, social media networks, such as Facebook and Twitter, automatically modify image files that are uploaded by users or copied from other websites (such as compression, resolution reduction and scaling) to fit the platform's needs. These modifications will change the cryptographic hash values too and thus void the verification by end user's browser extension. For this reason, reposting on social media platforms need alternative hash than traditional cryptographic hash. We believe *perceptual hash* [14] could be a plausible method to be used in the certification process.

Another related limitation is that the certification generated XMP sidecar file contains all the data needed for the endorsement process. Thus, it must be accompanied with the original news content when the news is reposted by other websites.

### B. Trust and Reputation for Small Publisher and Individual Blogger

The proposed fake news defense system gives big and well-known news organization the capability to endorse or certify true news. However, it does not help smaller organizations or individual to make their voice heard. If a news content is certified by an unknown news organization, or even by an individual blogger, few people will trust this certification, except the loyal followers of that certifying blogger.

In other words, the proposed fake news defense system in this paper is a top-down authoritative approach. This is not comprehensive enough. For the current Internet where many people consume news mainly from various social media networks, we also need to design a bottom-up "democratic" approach in news certification based on crowdsourcing.

### C. *Browser Extension Compatibility*

JavaScript to read the public key from the server's TLS certificate requires code that runs in the secure context of a browser extension (outside the webpage) to read the TLS certificate chain. A 3rd party library (PKI.js) provides many of the other cryptographic utilities. Currently, our prototype browser extension is developed for the Mozilla Firefox browser, but it is straightforward to develop a version that is compatible with other browsers such as Google Chrome and Safari.

### D. *Implementation and Support for Large Video*

The proposed fake news defense certification system is applicable to any multimedia news content. However, news that containing large video presents several challenges for our web browser extension. The main issue is related to the streaming process. If the certification is on the entire video file, the browser extension will not be able to verify the certification until the entire video file has been received. Our next research on this topic is to resolve certification for video streaming news content. One possible approach is to divide the video streaming into many smaller chunks, and we generate digital signature for every chunk of video data. In this way, the certification can be conducted as the video is continuously streamed. Another idea is to take advantage of video streaming technologies and video format to provide real-time certification verification by our browser extension code.

### E. *Support for Multi-version Web Page Publishing*

Some web publishing platforms can automatically serve multiple sizes of an image or video depending on the client browser, client Internet access speed, and device screen resolution. Furthermore, bandwidth-saving algorithms on a server or client device may also dynamically choose a suitable size or resolution of image or video for each individual end user. So-called "mobile friendly" modifications of the image will result in a different cryptographic hash checksum, and the digital signature will be invalid. A possible solution to resolve this challenge is that the server-side program will generate the XMP sidecar file for each version of an image or video file. For dynamically modified multimedia content, the server-side certification code needs to be run in real time on the server to generate the corresponding digital signature and associated sidecar file.

## VII. CONCLUSION

Our paper has demonstrated that by utilizing existing digital certificate infrastructure and digital signature technique, we can provide a reliable and easy-to-deploy system to pass the endorsement of a news organization on newsworthy multimedia content to end consumers. This design provides a new paradigm in fighting against fake news attack. We realized this design by using a sidecar file, that is easily created with our scripts, to contain the digital signature of the news content. In conjunction with our browser extension, the sidecar file is used to validate the image as they are loaded by Internet browser. The user is then quickly alerted to an altered or fake image with an easy to identify marker on the image. If the user does not have, or has disabled the browser extension, the sidecar file is completely ignored by the user's Internet browser. This design provides transparency to the end user and allows for a voluntary and graduate adaptation and deployment.

## ACKNOWLEDGMENT

This work was sponsored by the U.S. National Science Foundation (NSF) under Grant DGE-1915780.